%% file: ms.tex
\def\mode{1} % two-column with figures in the middle
\if 0\mode
    \documentclass[preprint,superscriptaddress,floatfix]{revtex4-2}
    \usepackage[left]{lineno}
    \linenumbers
\else
    \documentclass[aps,pra,twocolumn,superscriptaddress,floatfix]{revtex4-2}
\fi

\usepackage{placeins}
\usepackage{afterpage}
\usepackage{environ}
\usepackage{fullpage}
\usepackage{amssymb,amsmath}
\usepackage[utf8x]{inputenc}
\usepackage[T1]{fontenc}
\usepackage{siunitx}
\usepackage[version=3]{mhchem}
\usepackage[dvipsnames]{xcolor}
\usepackage{xfrac}
\newcommand\captionof[1]{\def\@captype{#1}\caption}

\usepackage{natbib}
\usepackage[unicode=true]{hyperref}
\hypersetup{breaklinks=true,
            colorlinks=false,
            pdfborder={0 0 0}}
\usepackage{graphicx}
\makeatletter
\def \figwidth{
    \if 0\mode
        0.5\textwidth
    \else
        \columnwidth
    \fi
}
\makeatother

\newcommand\Exp[1]{\ensuremath{\langle{#1}\rangle}}

\newcommand\tun{\,\hbar/t}
\newcommand\exch{\,\hbar/J}

\makeatletter
\newwrite\remember@figures
\AtBeginDocument{\InputIfFileExists{\jobname.dft}{}{}\immediate\openout\remember@figures=\jobname.dft}
\AtEndDocument{\immediate\closeout\remember@figures}
\NewEnviron{dfigure}[1]{%
    \immediate\write\remember@figures{%
        \noexpand\rememberfigure{#1}{\unexpanded\expandafter{\BODY}}%
    }%
}
\NewEnviron{dfigure*}[1]{%
    \immediate\write\remember@figures{%
        \noexpand\rememberfiguretc{#1}{\unexpanded\expandafter{\BODY}}%
    }%
}
\newcommand{\placefigure}[2][tp]{%
    \csname remembered@figure@#2\endcsname{#1}%
}
\newcommand{\rememberfigure}[2]{%
    \global\@namedef{remembered@figure@#1}##1{%
        \begin{figure}[##1]#2\end{figure}%
    }%
}
\newcommand{\rememberfiguretc}[2]{%
    \global\@namedef{remembered@figure@#1}##1{%
        \begin{figure*}[##1]#2\end{figure*}%
    }%
}
\makeatother

\begin{document}
\input{text_main}

\setcounter{figure}{0}
\renewcommand{\thefigure}{S\arabic{figure}}

\clearpage
\section*{Supplemental Material}
\input{text_supp}

\end{document}

%% file: text_main.tex
\title{Coupling a mobile hole to an antiferromagnetic spin background:\\ Transient dynamics of a magnetic polaron}

\newcommand{\harvard}{Department of Physics, Harvard University, 17 Oxford St., Cambridge, MA 02138, USA}
\newcommand{\tum}{Department of Physics and Institute for Advanced Study, Technical University of Munich, James-Franck-Str. 1, 85748 Garching, Germany}
\newcommand{\mcqst}{Munich Center for Quantum Science and Technology, Schellingstr. 4, 80799 M\"unchen, Germany}
\newcommand{\lmu}{Department of Physics and Arnold Sommerfeld Center for Theoretical Physics, Ludwig Maximilian University of Munich, Theresienstr. 37, 80333 M\"unchen, Germany}

\author{Geoffrey Ji}
\author{Muqing Xu}
\author{Lev Haldar Kendrick}
\author{Christie S. Chiu}
\author{Justus C. Brüggenjürgen}
\author{Daniel Greif}
\affiliation{\harvard}
\author{Annabelle Bohrdt}
\affiliation{\tum}
\affiliation{\mcqst}
\author{Fabian Grusdt}
\affiliation{\mcqst}
\affiliation{\lmu}
\author{Eugene Demler}
\author{Martin Lebrat}
\author{Markus Greiner}
\affiliation{\harvard}

\date{\today}

\begin{abstract}
Understanding the interplay between charge and spin and its effects on transport is a ubiquitous challenge in quantum many-body systems. In the Fermi-Hubbard model, this interplay is thought to give rise to magnetic polarons, whose dynamics may explain emergent properties of quantum materials such as high-temperature superconductivity. In this work, we use a cold-atom quantum simulator to directly observe the formation dynamics and subsequent spreading of individual magnetic polarons. Measuring the density- and spin-resolved evolution of a single hole in a 2D Hubbard insulator with short-range antiferromagnetic correlations reveals fast initial delocalization and a dressing of the spin background, indicating polaron formation. At long times, we find that dynamics are slowed down by the spin exchange time, and they are compatible with a polaronic model with strong density and spin coupling. Our work enables the study of out-of-equilibrium emergent phenomena in the Fermi-Hubbard model, one dopant at a time.
\end{abstract}

%\keywords{}

\maketitle

\section*{Introduction}

Interactions between charge carriers and magnetic excitations can drastically alter the transport properties of a many-body system. Prominent examples include the Kondo effect, colossal magnetoresistance and heavy-fermion materials, where electrical resistivity is strongly affected by electron scattering against magnetic impurities and localized spins.
In the two-dimensional Hubbard model, the competition between charge delocalization, related to the nearest-neighbor tunneling energy $t$, and antiferromagnetism, set by the spin exchange energy $J$, already results in intricate dynamics at the level of a single charge excitation. This iconic problem has attracted extensive theoretical attention \cite{brinkman_single-particle_1970, nagaev_ferromagnetic_1975, schmitt-rink_spectral_1988, kane_motion_1989, sachdev_hole_1989, liu_dynamical_1992, dagotto_correlated_1994, mishchenko_electron_2004, manousakis_string_2007} and can be reinterpreted as the creation and dispersion of a magnetic polaron, a charge carrier dressed by the magnetic background with renormalized properties. Because of their out-of-equilibrium character, knowing both transient and long-time dynamical properties of polarons is a fundamental step towards elucidating intriguing transport phenomena seen in strongly correlated materials.

Evidence for magnetic polarons in undoped cuprates has been seen in the form of a renormalized hole dispersion measured via photoemission experiments \cite{damascelli_angle-resolved_2003}. However, an understanding of the transient behavior of a single charge excitation is obscured in solid-state systems by the presence of phonons and inherently fast electron dynamics. Ultracold atoms offer a clean and tunable quantum simulation platform where quantum dynamics can be observed with a resolution finer than all relevant timescales \cite{cetina_ultrafast_2016}. Quantum gas microscopy, in particular, gives access to spin and density readout and manipulation at the single-site level, thereby enabling demonstrations of one-dimensional quantum walks \cite{weitenberg_single-spin_2011, preiss_strongly_2015} and, more recently, the study of spin-charge deconfinement in one-dimensional Hubbard chains \cite{vijayan_time-resolved_2020}. Previous dynamical studies of the doped 2D Fermi-Hubbard model have focused on hydrodynamic bulk transport \cite{anderson_conductivity_2019,brown_bad_2019, nichols_spin_2019}, and equilibrium signatures of polaronic behavior have been seen in small systems \cite{chiu_string_2019, koepsell_imaging_2019}.

Nonetheless, observing the competition between density and spin dynamics is experimentally challenging. It requires simultaneously achieving low temperatures for the spin order to dramatically influence transport, and large system sizes to probe the system's evolution over long spin exchange timescales.
In this work, we study the density- and spin-resolved dynamics of a single hole interacting with a two-dimensional antiferromagnetic Mott insulator of 400 sites (Fig.~\ref{fig:fig1}a). We first observe fast dressing of the spin environment over the tunneling timescale, where quantum interference between alternative hole paths plays a large role. This fast proces is followed by a slow delocalization of the hole and a slow relaxation of the spin background, which are key dynamical signatures of a magnetic polaron. We quantitatively validate our understanding of the spin dynamics by comparing them to a phenomenological model of the formation and departure of a spin polaron.

%TC:break main

\if 1\mode
\placefigure[!t]{f1}
\fi

\section*{Examining a single dopant}
We prepare the initial system of fermionic lithium-6 in a half-filled 2D square lattice with Hubbard parameters $t/\hbar=2\pi \times 744(12)\,\mathrm{Hz}$ and $U/t=8.72(28)$ ($t/J = 2.18(7)$) \cite{chiu_quantum_2018} at a temperature of $T/t=0.340(19)$ and with correlation length $\xi = 1.695(11)$ (expressing lengths in units of the lattice spacing $a = 569$ nm and setting $k_B = 1$ here and subsequently). Simultaneously, we project repulsive light from two digital micromirror devices (DMDs). The first DMD performs entropy redistribution and removes residual harmonic confinement over a rounded-square-shaped area of 31 sites in diameter \cite{supplement,mazurenko_cold-atom_2017}. The second DMD acts to prevent the occupation of selected sites while adiabatically loading the optical lattice, resulting in pinned holes (Fig.~\ref{fig:fig1}b). We release the pinned holes by shutting off the light illuminating the second DMD within $0.03 \tun$, then allow the system to evolve for a variable time $\tau$. Finally, we freeze the dynamics by rapidly increasing the lattice depth, and make a projective site-resolved measurement of either the parity-projected (singles) density or a single spin state by removing the other via a resonant spin-removal laser \cite{parsons_site-resolved_2016}.

%TC:break f3

\section*{Short-time density dynamics}

To obtain a density distribution of the hole location, we average the singles density distribution at a given time $\tau$. We then subtract it from an average background obtained without the hole. This process allows us to first cancel out density fluctuations in the form of doublon-hole pairs, which are imaged as empty sites, and, second, to remove residual systematic offsets in the particle density that may result from lattice inhomogeneities \cite{supplement}. For an enhanced data collection rate at times $\tau < \tun$, we simultaneously study four holes arranged in a seven-site-wide square pattern, instead of a single central hole (Fig.~\ref{fig:fig1}b). The density on the pinned sites initially ranges from 0.71(5) to 0.921(27) without significantly affecting their adjacent sites \cite{supplement}.

\if 1\mode
\placefigure[!t]{f2}
\fi

The resultant hole density distributions at short times are shown in Fig.~\ref{fig:density}a. Within half a tunneling period, the hole tunnels to the four neighboring sites ($\tau = 0.47(1)\tun$); its subsequent propagation retains clear coherent features such as the oscillation of the hole density on the sites adjacent to the origin (maximal at $\tau = 0.47(2)\tun$), which sets it apart from a classical diffusion process.

We can gain microscopic insights on the role of spin by plotting the hole densities on the central site and its diagonal sites, as shown in Fig.~\ref{fig:density}b, at times shorter than the spin superexchange timescale $\hbar/J = 2.18(7) \hbar/t$.
Within two tunneling periods, hole motion is affected by the presence or absence of quantum interference effects between the two paths leading to the same corner site but possibly distinguishable spin backgrounds.
To quantify this effect, we show the predictions of three models that feature different magnetic phases: a noninteracting quantum walk, equivalent to the propagation of a hole in a spin-polarized background; a quantum Monte Carlo (QMC) simulation on a disordered spin background ($T = \infty$, $J = 0$) \cite{carlstrom_quantum_2016,kanasz-nagy_quantum_2017}; and a time-dependent density matrix renormalization group (TD-DMRG) simulation of the $t-J$ model with $t/J = 2$ on a $18\times4$ system, initially in the antiferromagnetic ground state \cite{bohrdt_dynamical_2020}.
All three models quantitatively describe the experimental results during the first tunneling event up to $\tau \approx 0.5 \tun$; a detailed comparison is presented in \cite{supplement}. Afterwards, the spinful simulations predict a revival of the central density around $\tau \approx 1.2 \tun$, though finite-size effects magnify this revival in the TD-DMRG simulation. The amplitude of the oscillation in the diagonal density is directly related to the indistinguishability of spin backgrounds after two tunneling events ending at a given site, as sketched in Figs.~\ref{fig:density}c and d: Quantum interference is maximal in a ferromagnet (free quantum walk), reduced in an antiferromagnet (ground-state TD-DMRG), and between these two extremes in a paramagnet (infinite-temperature QMC). The suppressed diagonal density seen experimentally at $\tau \approx 1 \tun$ therefore hints at the quantum statistical role of the antiferromagnetic background in the hole dynamics over short timescales.

\section*{Long-time hole delocalization}

The previous experimental data focus on times smaller than the spin exchange time $\hbar/J$, where the hole motion is essentially driven by direct tunneling. Investigating dynamics at longer times is challenging, as it requires resolving hole densities inversely proportional to the square of its distance to the origin. To mitigate statistical fluctuations in the hole density at large distances, we fit the experimental data in an analysis region containing most of the hole density to a two-dimensional Gaussian distribution that empirically match the hole distribution at longer times \cite{supplement}. The delocalization dynamics is then quantitatively described by the root-mean-squared (RMS) hole distance:
\[d_\text{RMS} = \sqrt{\sum_{d_x, d_y}{(d_x^2 + d_y^2) \rho_\textbf{d}} \bigg/ \sum_{d_x, d_y}{\rho_\textbf{d}}},\]
where $\rho_\textbf{d}$ is the fitted hole density at coordinate $\textbf{d} = (d_x, d_y)$ relative to the initial hole position.

\if 1\mode
\placefigure[!t]{f3}
\fi

We plot the RMS distance of the fitted distribution in Fig.~\ref{fig:distance}. At times $\tau < 1\tun$, we observe a fast linear growth characteristic of a ballistic expansion. The data match the analytical expression for a noninteracting quantum walk, $d_\text{RMS} = v \tau$ with velocity $v = 2 t/ \hbar$, confirming that the spin background can be effectively neglected during the first tunneling event. The delocalization then clearly slows after $\tau = 1\tun$; the hole eventually leaves our analysis region of 11 sites in radius (Fig.~\ref{fig:fig1}b) at times greater than $\tau = 10\tun$ \cite{supplement}.

We can qualitatively capture this slowdown with an analytical model depending only on $t$, mapping hole motion on a spinful background to a free quantum walk on a Bethe lattice \cite{golez_mechanism_2014,kanasz-nagy_quantum_2017}. The model predicts a crossover from ballistic to diffusive behavior resulting from the absence of quantum interference between paths leading to the same hole position but different spin backgrounds \cite{carlstrom_quantum_2016}. Its RMS distance asymptotically follows a square-root law $d_\text{RMS} = \sqrt{4 D \tau}$ with a diffusion constant $D \approx 1.37 D_\text{MIR} = 1.37 t a^2 / \hbar$ close to the Mott-Ioffe-Regel (MIR) limit, which was experimentally shown to be a lower bound for diffusion at larger doping \cite{brown_bad_2019}. A similar ballistic-to-diffusive crossover occurs in spin-1/2 QMC simulations at infinite temperature, with a diffusion constant increased by less than 10\% compared to the Bethe lattice case with effective infinite spin of the Bethe lattice, due to enhanced quantum interference \cite{kanasz-nagy_quantum_2017}.

Though the Bethe-lattice model qualitatively predicts the slowdown of the hole, its RMS distance exceeds that of the experiment at long times (with a chi-square value $\chi^2 = 47$ on the six experimental times $\tau > \tun$ corresponding to a $p$-value of $p = 2 \times 10^{-8}$). This deviation challenges the assumption of tunneling being the only relevant energy scale. To directly verify the role of the superexchange on hole motion, we use a Feshbach resonance to double the on-site interaction to $U/t=17.2(6)$ ($t/J=4.30(15)$) at fixed tunneling. We thereby halve the superexchange energy $J$ while slightly decreasing the temperature $T/t=0.241(18)$ and increasing the correlation length $\xi = 2.48(25)$. The experimental RMS distance shown in Fig.~\ref{fig:distance} agrees with the previous data at short times, apart from small initial deviations due to a larger hole preparation infidelity \cite{supplement}. A linear fit for data at $\tau > 0.8 \exch$ indicates a reduction of the long-time hole velocity for smaller $J$ from a value of $0.40(10)\,a/(\hbar/t)$ at $U/t = 8.72(28)$ down to $0.15(17)\,a/(\hbar/t)$ at $U/t=17.2(6)$. These results are consistent with the expected decrease of the quasiparticle bandwidth with spin exchange in the $t-J$ model in the strong-coupling limit $t \gg J$ \cite{sachdev_hole_1989}, and are quantitatively compatible with the numerical predictions in \cite{dagotto_strongly_1990,martinez_spin_1991,liu_dynamical_1992,leung_dynamical_1995,grusdt_microscopic_2019}.
This agreement suggests that at long times the hole becomes dressed by the magnetic background and forms a magnetic polaron.

\if 1\mode
\placefigure[!htbp]{f4}
\fi

%TC:break f4

\section*{Spin dynamics}
The two-stage dynamics of the hole motion are also visible in the spin correlations, whose fast evolution away from equilibrium and slow return to it can be interpreted as the formation and departure of a polaron, respectively.
We measure the sign-corrected spin correlation function $C_{\textbf{r}}(\textbf{d}) = (-1)^{d_x+d_y}4 ( \Exp{S^z_\textbf{r} S^z_{\textbf{r}+\textbf{d}}} - \Exp{S^z_\textbf{r}} \Exp{S^z_{\textbf{r}+\textbf{d}}} )$ in an independent data set starting with a single hole at the center \cite{supplement}. We plot a map of the correlations from the initial hole location $C_\textbf{0}(\textbf{d})$ (as the hole becomes indistinguishable from hole fluctuations once it starts moving) and averaged over all spatial symmetries for select times, see Fig.~\ref{fig:correlation_reversal}a. At $\tau = 0$, these correlations are vanishing because of the presence of the hole; within one tunneling time, the hole hops to a neighboring site and the correlations become those of the exchanged neighboring spin (Fig.~\ref{fig:correlation_reversal}b). This swapping of correlations results in a reversal of the global antiferromagnetic correlation pattern (light brown color in Fig.~\ref{fig:correlation_reversal}a): For instance, the negative sign of the diagonal correlation $C_\textbf{0}(\textbf{d} = (1, 1))$ at $\tau = 0.467(8) \tun$ is at odds with its value at long times, $\tau = 23.4(4) \tun$.
Here, the reversal of the antiferromagnetic pattern extends up to 3 sites away from the center and is a dynamical analog to the short-range polaronic behavior seen at equilibrium in Ref.~\cite{koepsell_imaging_2019}. In contrast, we find that the nearest-neighbor correlator does not exhibit the same sign-change effect (light green squares at $\tau = 0.467(8) \tun$) as it results from a mixture of diagonal correlations weakened by the presence of the hole before the initial quench.

The swapping mechanism described in Fig.~\ref{fig:correlation_reversal}b suggests a practical way to quantitatively predict the evolution of spin correlations resulting from fast hole dynamics. To do so, we take experimental pictures at $\tau=0$ and displace spins according to the distribution of hole trajectories predicted by the analytical Bethe-lattice model \cite{supplement}. In Fig.~\ref{fig:correlation_map}a, we compare the time evolution of the next-nearest correlations to the center $C_\textbf{0}(|\textbf{d}| = \sqrt{2})$ with the predicted evolution based on this model, shown as a purple-gray band. It agrees with the experimental data at short times only, indicating that local correlation swapping is an accurate picture in that regime.

A more complete model for the hole dynamics is one that considers the energy cost of correlation swapping, which, at first, binds the hole to around its initial location \cite{bulaevskii_new_1968}. Spin exchange can then enable the restoration of the disrupted spin background, leading to a polaron delocalizing with a long-time velocity set by the superexchange energy. An example of such a model is presented in Ref.~\cite{grusdt_microscopic_2019}, where the polaron is described as a composite object formed by a holon and spinon connected by a string of displaced correlations. We include the spinon dynamics by shifting the experimental pictures used in our model (and, therefore, the effective hole origin) according to a ballistically propagating spinon \cite{supplement}. Including these dynamics results in an accurate prediction of the long-time behavior of the spin correlations, qualitatively capturing their slow return to equilibrium (blue band in Fig.~\ref{fig:correlation_map}a). Spin correlations relax more slowly than the local density (inset of Fig.~\ref{fig:correlation_map}a) -- strikingly, equilibration in the spin sector has not occurred even for the last measured time of $\tau = 23.4(4) \tun$ although the hole itself has left the system.
% These slow dynamics suggest a new spin relaxation timescale appears as a consequence of the charge dynamics.

\if 1\mode
\placefigure[!hbtp]{f5}
\fi

The polaron model can also be benchmarked with spin correlations measured across the entire Mott-insulating region. We show in Fig.~\ref{fig:correlation_map}b the nearest-neighbor correlations $C_{\textbf{r}}(|\textbf{d}| = 1)$ averaged according to the bond distance to the center $|\textbf{r}|$ (full data shown in \cite{supplement}). Strongly depleted correlations are initially observed around the center due to the hole, before spreading towards larger distances as the system relaxes. Since nearest-neighbor correlations are directly proportional to the local magnetic energy in the $t-J$ approximation, this propagation indicates that the hole imparts its kinetic energy to the entire spin background while leaving the systems.

Including spinon-holon dynamics (solid lines in Fig.~\ref{fig:correlation_map}b) compared to a bare Bethe-lattice model (dashed lines) leads to good agreement with the experimental data even at longer times, confirming the role of spin exchange in the long-time dynamics. However, theory underestimates nearest-neighbor correlations at short distances and short times $\tau = 0.94$ and $2.35 \tun$, perhaps as a result of spin relaxation mechanisms not captured in the model such as magnon emission.
Furthermore, we note that the model overestimates the RMS distance of the hole measured experimentally \cite{supplement}, possibly because it neglects spin fluctuations that effectively lead to a disordered magnetic background and inhibit hole motion.

%TC:break conclusion

In this work, we demonstrate the intricate linkage between charge and spin dynamics in the two-dimensional Fermi-Hubbard model and explore the formation and motion of magnetic polarons. The size, coherence and control of this cold-atom quantum simulator are key to benchmarking nonequilibrium theories of magnetic polaron formation and numerical techniques away from the linear-response regime. Our work could be extended to investigate the interaction between multiple deterministically created polarons, and to study how magnetism facilitates this interaction. This extension may enable the microscopic observation of $d$-wave Cooper pairs, and of emergent phases such as stripe phases, pseudogap phases and ultimately $d$-wave superconducting phases.

\begin{dfigure}{f1}
    \centering
    \noindent
    \includegraphics[width=\figwidth]{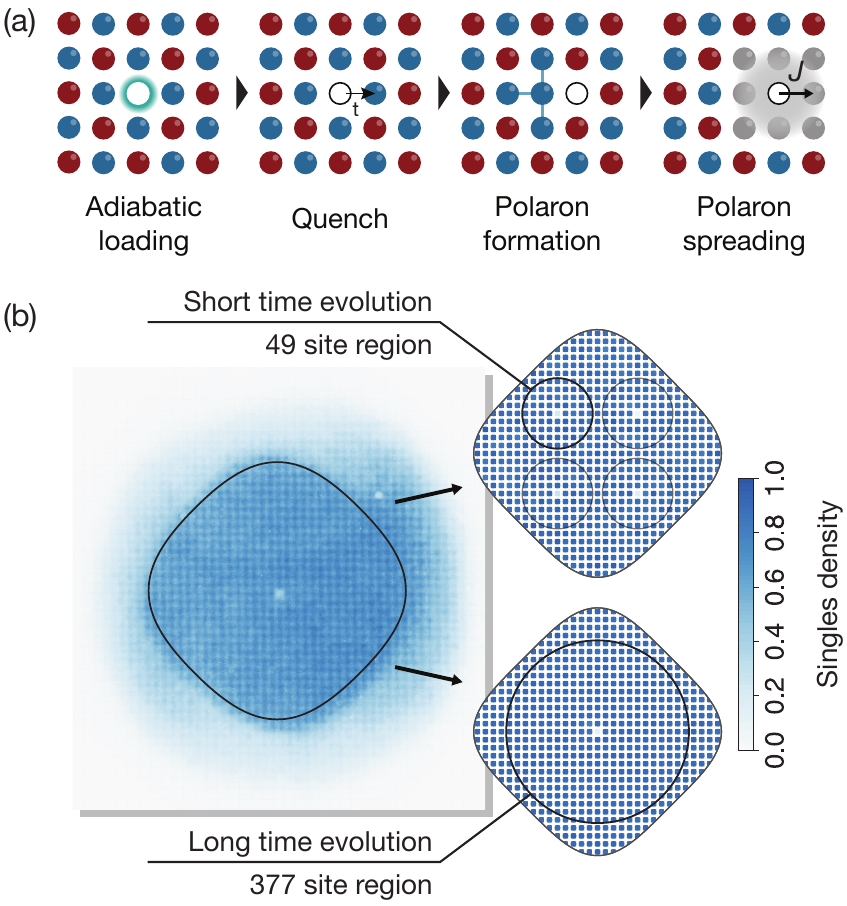}
    \caption{\textbf{Probing single-hole dynamics with a quantum gas microscope.}
    (\textbf{a}) We initialize a hole in the system by adiabatically loading atoms into the lattice with a localized repulsive potential. The potential is abruptly removed, allowing the hole to delocalize at the bare tunneling rate $t$ and displace spins along its path that result in a local reversal of spin correlations compared to equilibrium. The hole is therefore locally dressed by the spin background, forming a magnetic polaron with a velocity ultimately determined by $J$.
    (\textbf{b}) (Left) Average of experimental images before hole release for one prepared hole. The apparent additional hole in the upper right is caused by defects in the imaging path. (Right) Average density of the atomic system before hole release, for one hole and four holes (to enhance the data collection rate at short evolution times). All analysis is done inside the denoted regions.}
    \label{fig:fig1}
\end{dfigure}

\begin{dfigure}{f2}
    \centering
    \noindent
    \includegraphics[width=\figwidth]{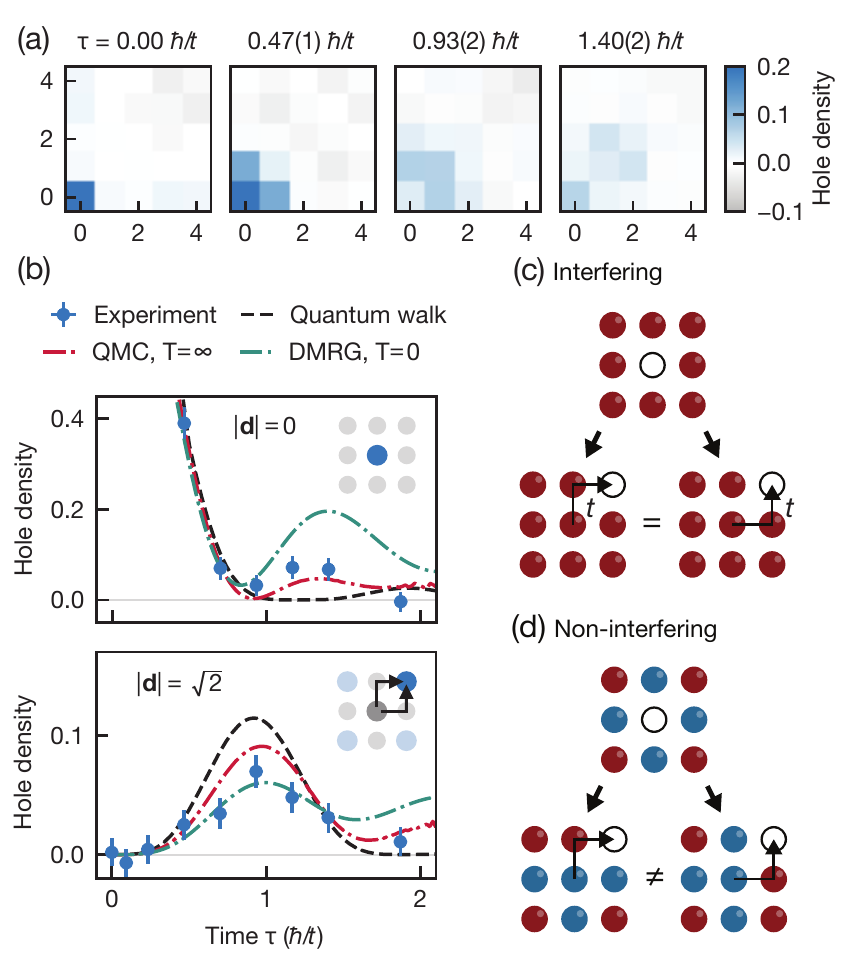}
    \caption{\textbf{Effect of quantum interference on short-time density dynamics.}
    (\textbf{a}) Symmetrized average hole densities at short times. The hole initially delocalizes coherently, as indicated by density oscillations on sites surrounding the origin.
    (\textbf{b}) Short-time evolution of the hole density at the center (distance $|\textbf{d}| = 0$) and on the diagonally adjacent sites ($|\textbf{d}| = \sqrt{2}$). We compare experimental data to models featuring a ferromagnetic (noninteracting quantum walk, dashed black line), disordered (noninteracting, spin-1/2 QMC simulation at $T = \infty$, dash-dotted red line) or antiferromagnetic spin background (ground-state DMRG simulation at $t/J = 2$, dash-dotted green line). The diagonal density shows a decreased oscillation amplitude compared to the quantum walk and QMC simulations. This can be interpreted as a manifestation of quantum interference between two nonequivalent hole paths, maximal for a polarized spin background (\textbf{c}) and reduced for an antiferromagnet (\textbf{d}). 
    Here and in the following, error bars indicate a one-sigma statistical uncertainty in the plotted values.
    }
    \label{fig:density}
\end{dfigure}

\begin{dfigure}{f3}
    \centering
    \noindent
    \includegraphics[width=\figwidth]{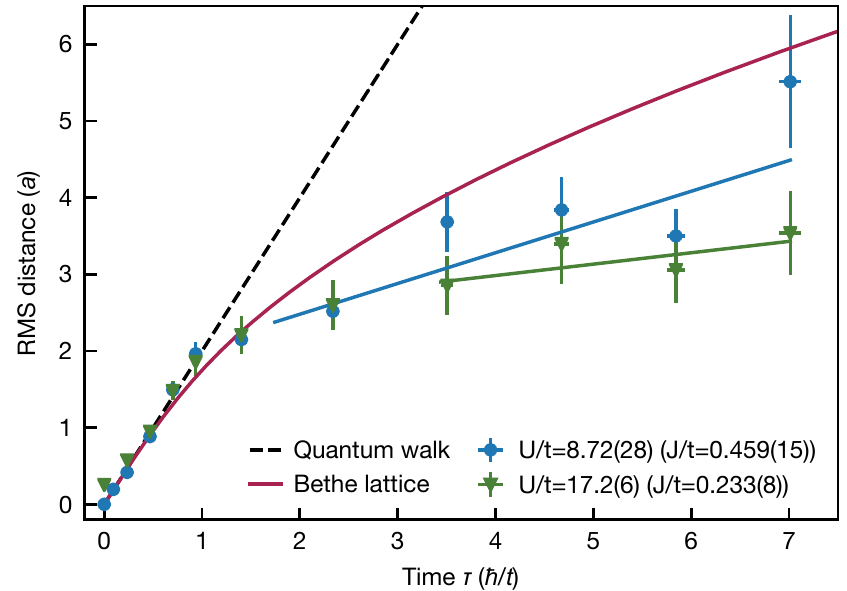}
    \caption{\textbf{Long-time slowdown of the hole propagation due to superexchange.}
    Root-mean-squared distance from the origin, obtained from a Gaussian fit to the 2D hole density distribution. The initial linear increase indicates a ballistic expansion compatible with a noninteracting quantum walk (dashed black line). The hole slows down around $\tau = 1 \tun$, an effect qualitatively predicted by an analytic model based on a free quantum walk in a Bethe lattice (solid blue line). At later times, the hole is more confined than the Bethe-lattice prediction. Increasing the interaction energy decreases the superexchange and the long-time hole velocity, from $0.40(10)\,a/(\hbar/t)$ and  at $U/t = 8.72(28)$ to $0.15(17)\,a/(\hbar/t)$ at $U/t = 17.2(6)$ as obtained from linear fits at $\tau \ge 0.8 \exch$.}
    \label{fig:distance}
\end{dfigure}

\begin{dfigure}{f4}
    \centering
    \noindent
    \includegraphics[width=\figwidth]{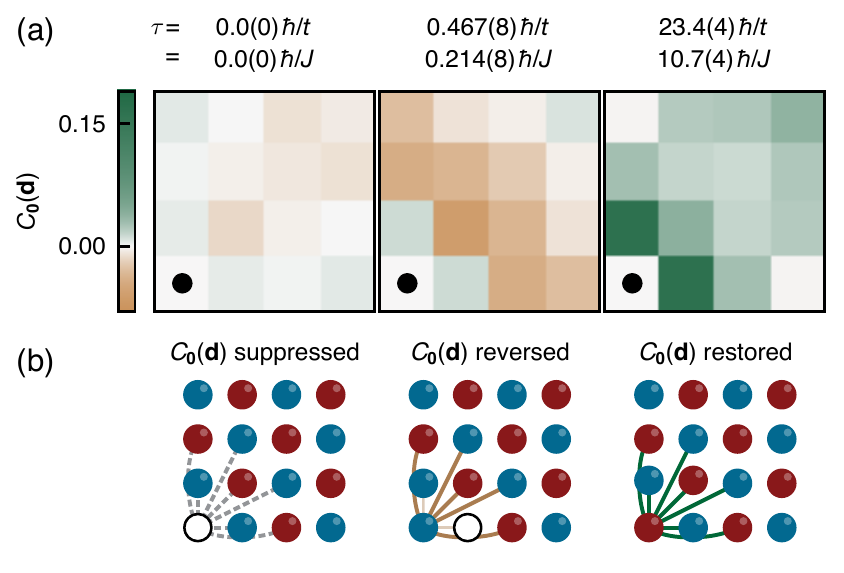}
    \caption{\textbf{Reversal and recovery of antiferromagnetic correlations.} (\textbf{a}) Sign-corrected spin correlations $C_{\mathbf{0}}(\mathbf{d})$ from the initial hole location (black circle) at select times, symmetrized across reflections. Within one tunneling time we observe antiferromagnetic correlations but with the opposite sign (light brown) apart from the nearest-neighbor correlator. The correct AFM pattern (green) then slowly restores itself, although the correlations have not fully equilibrated even at our longest measured times ($\tau = 23.4(4)\tun = 10.7(4)\exch$; \cite{supplement}) (\textbf{b}) The emergence of short-time antiferromagnetic correlations with the opposite sign can be understood in a classical Néel state. Initially, the hole swaps a neighboring spin to the origin, producing reversed correlations at the origin site. At later times, relaxation processes restore the correct antiferromagnetism.}
    \label{fig:correlation_reversal}
\end{dfigure}

\begin{dfigure}{f5}
    \centering
    \noindent
    \includegraphics[width=\figwidth]{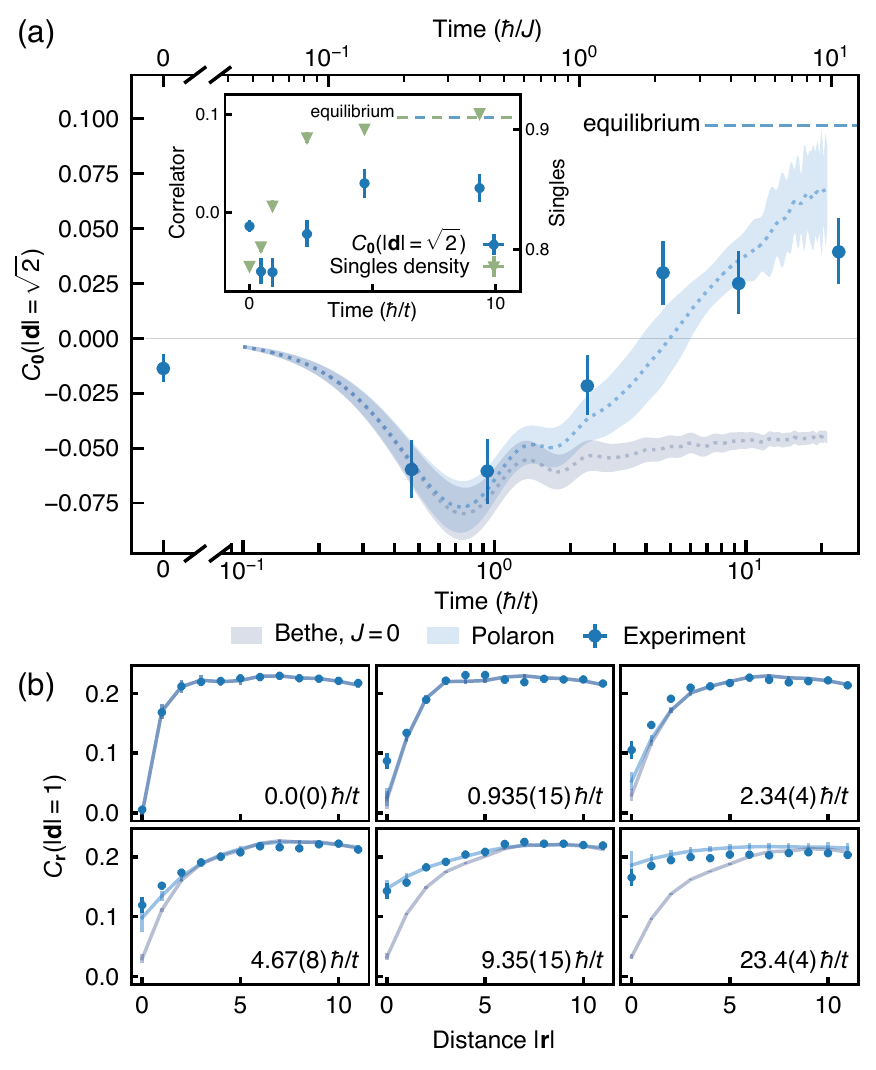}
    \caption{\textbf{Comparison with a polaron model.} (\textbf{a}). Sign-corrected spin correlations between the initial hole location and its diagonal neighbors ($C_{\mathbf{0}}(|\mathbf{d}|=\sqrt{2})$) as a function of time. The short-time dynamics are captured by a Bethe-lattice model only considering charge motion (purple-gray band), while spin exchange must be taken explicitly into account (here as a polaron model, blue band) to qualitatively describe the long-time relaxation of the correlations towards their equilibrium values (blue dashed line). This correlation relaxation is slower than the density relaxation averaged on a $3 \times 3$ area around the center (inset). (\textbf{b}) Nearest-neighbor correlations as a function of distance to the center $C_{\mathbf{r}}(|\mathbf{d}|=1)$. The slow spreading of the depleted correlations towards the system edges is best captured by the polaron model (blue) as opposed to the Bethe-lattice model (purple-gray). Neither model quantitatively captures the short-time dynamics at the center ($|\mathbf{r}|=0$).}
    \label{fig:correlation_map}
\end{dfigure}

\section*{Acknowledgments}

We thank E. Altman, J. Carlström, M. Kanász-Nagy, J. Léonard, M. Knap, A. Lukin, E. Manousakis, Z.-X. Shen and W. Zwerger for insightful discussions.
% \textbf{Funding:}
We acknowledge support from the Gordon and Betty Moore Foundation Grant No. 6791; NSF Grants No. PHY-1734011, No. OAC-1934598, and No. OAC-1934714; ONR Grants No. W911NF-11-1-0400 and No. N00014-18-1-2863; ARO Grant No. W911NF-20-1-0163; ARO/AFOSR/ONR DURIP; the U.S. Department of Defense through the NDSEG program (G.J.), the NSF GRFP (L.H.K. and C.S.C.), the Deutsche Forschungsgemeinschaft (DFG, German Research Foundation) under Germany's Excellence Strategy -- EXC-2111 -- 390814868 (A.B. and F.G.), the Swiss National Science Foundation and the Max Planck/Harvard Research Center for Quantum Optics (M.L.).

%%%%%%%%%%%%%%%%%%%%%%%%%%%%%%%%%%%%%%%%%%%%%%%%%%%%%%%%%%%%%%%%%%%%%%%%%%%%%%

%% file: text_supp.tex
\subsection*{Experimental Stability}
To ensure consistent lattice position, atom number, and alignment of the hole DMD potential to the lattice, we perform active feedback on the experiment. The feedback corrects for drifts on the timescale of hours. We maintain a running average of the mean atom position and atom number and feedback on the angle of the lattice in-coupling mirrors and the final evaporation power. This ensures a total atom number of $1094 \pm 29$ for the short-time $U/t = 8.7(3)$ dataset over 1099 realizations ($99-102$ per evolution time), $1098 \pm 22$ for the long-time $U/t = 8.7(3)$ dataset over 7590 realizations ($754-762$ per evolution time), $919 \pm 27$ for the $U/t = 17.2(6)$ dataset over 2873 realizations ($190-194$ per evolution time), and $1099 \pm 25$ for the correlation dataset over 13255 realizations ($585-764$ per evolution time and per spin or no spin removal). For each dataset, an experimental average is obtained by changing the evolution time between two consecutive experimental realizations and looping through this sequence of evolution times repeatedly. Interleaving data acquisition therefore allows us to average residual, slowly varying systematic offsets in a similar way for different evolution times.

To ensure alignment of the hole DMD potential, we periodically display a superlattice potential on the DMD and measure the positional phase of the atomic distribution. We then offset the position of the DMD pattern accordingly.

\subsection*{Uniformity of the Potential Background}

\if 1\mode
\placefigure[!htbp]{uniformity}
\fi

The overall Gaussian envelope of the two radial lattice beams results in an overall harmonic confinement that is corrected by projecting an anti-harmonic potential with a Digital Micromirror Device.

We ensure that the residual potential curvature is negligible and the anti-confining potential does not introduce additional disorder over short distances. To do so, we load the central region with atoms in the compressible, metallic regime at an average singles density of $\bar{\rho}_s = 0.46(6)$ over 90 realizations (Fig.~\ref{fig:uniformity}a). We then convert the density to chemical potential $\mu(r)$ in the local density approximation, using the equation of state of the Fermi-Hubbard model numerically computed at $U/t = 8.7(3)$ using non-linear cluster expansion (NLCE) \cite{khatami_thermodynamics_2011}, and extrapolated down to experimental temperatures $T/t = 0.340(19)$. Assuming that the system is in thermal equilibrium at a global chemical potential $\bar{\mu}$, we can infer the local on-site potential $V(r) = \bar{\mu} - \mu(r)$. The variations of density correspond to a site-to-site RMS deviation of $V(r)$ of about $0.4 t$ across the filled region, and are small compared to both bandwidth $8 t$ and on-site interaction energy $U = 8.72(28)$ and $17.2(6) t$. This energy scale furthermore remains one order of magnitude smaller than the typical disorder strength where many-body localization is experimentally relevant in related 2D systems \cite{choi_exploring_2016}.

\if 1\mode
\placefigure[!htbp]{initial_density}
\fi

To empirically probe the effect of residual on-site energy offsets on the short-time hole motion, we prepare about 20 holes on a square grid with a spacing of 4 lattice sites (Fig.~\ref{fig:uniformity}b). We then image the system after a release time of $\tau = 0.7(1) \hbar/t$, which is about the time where the hole density is expected to be maximal on the four sites directly adjacent to the initial hole locations. We do not observe any long-range density variation from hole to hole other than inhomogeneities resulting from the decrease of the initial hole fidelity away from the center. Among lattice sites adjacent to the same hole, we observe a two-sigma density difference between sites highlighted by red links in Fig.~\ref{fig:uniformity}b after averaging over 164 realizations. They represent a little more than 5\% of all pairs of sites adjacent to a hole with an initial fidelity above 25\%, which is comparable to the proportion of outliers expected for a two-sigma confidence interval. We therefore do not conclude about the existence of anisotropies in hole tunneling given our statistical resolution.

\subsection*{Hole Preparation}

We initially create holes inside the two-dimensional Mott-insulating region by exposing one or several lattice sites to localized blue-detuned light created by a Digital Micromirror Device (DMD). This initial state is obtained by first ramping up the DMD potential in $30$ ms and then ramping up the lattice potentials in $60$ ms to prepare a system in equilibrium with the presence of the repulsive potential. The choice of the DMD potential height is a tradeoff between high hole probability on the target sites and unwanted atom depletion of neighboring sites due to an imperfect point spread function. Such imperfections prevent us from reliably prepare a single hole at higher temperatures due to the higher compressibility of the Fermi-Hubbard model away from the Mott phase, which is why we restrain ourselves to temperatures close to the antiferromagnetic transition in this work.

We show in Fig.~\ref{fig:initial_density} the initial hole density for the different datasets used in this work. To measure short-time site densities (Fig.~1), we decrease the initial density of four sites to hole fidelities ranging from 0.64(5) to 0.802(34), with variations caused by the non-uniform illumination of the DMD. To combine the data from each hole, we normalize each hole density distribution separately after background subtraction (see below) such that at $t=0$ the hole density at the pinned site is unity. The four hole density distributions are then averaged together.

\if 1\mode
\placefigure[!p]{density_full}
\fi

To measure the time evolution of the average hole distance (Fig.~3) and the correlations of the spin background (Figs.~4 and 5), we use a stronger pinning potential that leads to a higher hole fidelity at the origin and an increase of the average hole density on the neighboring sites. This depletion is associated to an excess hole number of about 0.15 in Fig.~\ref{fig:initial_density}b, 0.25 in Fig.~\ref{fig:initial_density}c and 0.3 in Fig.~\ref{fig:initial_density}d. This additional density away from the central site leads at $U/t = 17.2(6)$ to an increased RMS distance of 0.26(4) as shown in Fig.~3. This hole excess is however quickly reduced during the initial ballistic expansion as it is diluted over an area growing as the square of the propagation time: at $\tau = 0.47(2) \tun$, the fitted distance already agrees within error bars with the prediction for a free quantum walk.

\subsection*{Error on the Evolution Time}
We obtain a normalized evolution time by multiplying the experimental elapsed time by the tunneling energy in units of $\hbar$ obtained by lattice modulation spectroscopy \cite{parsons_site-resolved_2016}. The stated error includes the error in the spectroscopic measurement and the systematic error on the elapsed time as a result of the finite duration of the DMD shutoff time. We neglect error arising from the finite time of the lattice freezing ramp. We find that the hole distributions before and after releasing the hole are consistent, indicating that the freeze negligibly affects the hole distribution.

\if 1\mode
\placefigure[!htbp]{fit_total_in_mask}
\fi

\subsection*{Background Subtraction}
To determine the location of the prepared hole, we must measure and remove the equilibrium contribution from doublon-hole pairs due to quantum fluctuations and thermal effects. By loading the system with no hole potential and immediately imaging the system, we measure the equilibrium singles density where no hole is present $\rho^s_\text{eq} = 1 - (\rho^h+\rho^d)_\text{eq}$, where $\rho^h$ ($\rho^d$) is the average probability of an empty (doubly-occupied) site. The image is then convolved with a $3 \times 3$ kernel with a normalized weight of $0.15$ on the nearest-neighbor sites to remove high-frequency statistical noise. The hole density during hole propagation is estimated by subtracting the imaged singles density to the equilibrium one, $\rho = \rho^s_\text{eq} - \rho^s$. This subtraction process can cause statistical fluctuations to result in negative hole densities, which are nonetheless included in the Gaussian fit to estimate the average hole distance.  

In principle the effective background may change at different evolution times due to heating. We measure the heating rate to be $0.0009(3)\,\mathrm{t/ms}$, which has a negligible effect on the density even at the longest evolution time of $5\,\mathrm{ms}$.

\if 1\mode
\placefigure[!p]{fit_red_chisq}
\fi

\subsection*{Distance Fits}

To fit the hole density distribution, we first truncate the experimental data to a circular window of 4 or 11 sites in radius around the initial hole location, containing 49 or 377 sites, respectively. The smaller window is used for the low $U/t$ dataset where four holes are prepared, to avoid including the effect of the neighboring holes. The window contains most of the integrated hole density over the experimental timescales, as plotted in Fig.~\ref{fig:fit_total_in_mask}.

We fit to the experimental distribution a discrete two-dimensional Gaussian probability density function, where the probability on each site is equal to the integral of a continuous Gaussian function over the 1-site square corresponding to each lattice site. The discrete density distribution is centered on the initial hole location and normalized to the initial hole fidelity, and we use the width of the continuous Gaussian distribution as the only fit parameter. We numerically compute the RMS value of this distribution, and propagate the error from the fit assuming normally distributed errors.

We plot the fits as well as the reduced chi-squared test statistic versus the continuous fit parameter in Fig.~\ref{fig:fit_red_chisq}. We observe good quantitative agreement with the Gaussian functional form. We note that at $\tau t/\hbar=7.01$ for the high $U/t$ data we fit to a local minimum; the global minimum at large radius is a result of the low total density at that time. We believe this effect is a statistical fluctuation as the later data point at $\tau t/\hbar=9.35$ does not exhibit this issue. We do not show the fitted RMS at $\tau t/\hbar=9.35$ in the main text as the fit at low $U/t$ exhibits sensitivity to the exact window used. For the total density, we also plot the value at $\tau t/\hbar=23.4$, where we find that for both $U/t$ values the value is consistent with $0$, i.e. the hole having left the window. As a result, we do not fit the hole density distribution here.

\subsection*{Agreement of Gaussian Model}

\if 1\mode
\placefigure[!htbp]{fit_investigation}
\fi

For a hole undergoing a diffusive random walk process, the spatial hole probability distribution is expected to be be Gaussian. This distribution is a result of the central limit theorem and the fact that hole motion at each time interval is independent. However, if the hole propagation is not a memoryless process (therefore motion at each time interval may be correlated), we may not expect a Gaussian distribution. As a result, we examine the suitability of the Gaussian fit on both the Bethe-lattice model and the TD-DMRG data by comparing the fitted RMS with the true RMS, see Fig.~\ref{fig:fit_investigation}a.

The fit has excellent agreement with the Bethe-lattice model, especially at longer times. We believe this agreement is because hole motion in the Bethe lattice only displays interference effects when it retraces its path, which becomes increasingly unlikely for longer lengths. As a result, the hole motion appears diffusive and with a Gaussian shape at longer times.

The fit does not agree as well with the TD-DMRG data with system size $18 \times 4$, which has been integrated along the short dimension with periodic boundary conditions to alleviate finite-size effects. The TD-DMRG data has long-lived coherent oscillations along the 4-site rung including and surrounding the initial hole location, leading to an integrated distribution peaked at the center. Excluding the central rung from the fit results in a quantitatively accurate RMS. When excluding the central point from the Bethe and experimental fits, we see minimal deviation at longer times, further supporting the conclusion that importance of the center point is a finite-size effect. The fit of the TD-DMRG without central rung fails at the three smallest times as the central sites represents a substantial part of the probability density; these three points are not shown in Fig.~\ref{fig:fit_investigation}a.

We also investigate changing the window in which we fit the hole density distribution to, see Fig.~\ref{fig:fit_investigation}b and c. We see minimal deviation on the fitted RMS distance, with the exception of the last data point for the lower interaction energy data. There, the agreement is only statistical. We note that the two largest window sizes agree, however.

\if 1\mode
\placefigure[!htbp]{bondmap}
\fi

\subsection*{Quantum Walk, Bethe Lattice, QMC, and TD-DMRG Calculations}

The free quantum walk is calculated from the analytic formula in \cite{hartmann_dynamics_2004}, scaled to a two-dimensional system:
\begin{equation} \label{density}
    \rho_{i,j}(\tau) = |\mathcal{J}_{i}(2 \tau t)\mathcal{J}_{j}(2 \tau t)|^2,
\end{equation}
where $\mathcal{J}$ is the Bessel function of the first kind and $i$, $j$ are the site coordinates relative to the initial location.

The quantum walk on the Bethe lattice models the fact that the presence of the spin background means that different hole paths to the same lattice site lead to different quantum states, as the spin background is rearranged in different ways. This distinguishability is different from the case of a single-particle free quantum walk where all paths leading to the same lattice site interfere with one another. We can model this effect by instead considering the hole as moving coherently on a fractal Bethe graph, where hole hopping always leads to a new node unless the hole is retracing (going back on) its existing path. A single real-space lattice site maps to infinitely many nodes on this graph, as there are an infinite number of paths that reach the same lattice site. We perform this mapping classically, by summing the probability of all nodes that correspond to a given lattice site. This results in motion that appears diffusive at long times.

Hole propagation in the Bethe lattice and in a paramagnetic spin-\sfrac12 environment are calculated as in \cite{kanasz-nagy_quantum_2017}. In the latter case, we perform QMC in a $40\times40$ system with 100 runs and $1\times10^{7}$ paths, sampling both the forward and backward paths. We use the code provided at: \href{https://github.com/MartonKN/Dynamical-spin-correlations-at-infinite-temperature}{https://github.com/MartonKN/Dynamical-spin-correlations-at-infinite-temperature}.

The DMRG simulations are performed using the TeNPy package \cite{hauschild2018} on a $18\times4$ system with periodic boundary conditions. The time evolution is performed using the matrix product operator based $W^{II}$ method \cite{Kjaell2012,Zaletel2015,Gohlke2017}.
The TD-DMRG simulation has finite-size effects due to the short dimension being only 4 lattice sites. To ameliorate these effects when calculating the RMS distance, we first integrate the hole density distribution along the short dimension. This integration will result in the same distribution as a system with a larger short dimension as long as the configurations do not add coherently. We then calculate the RMS distance along the long direction, and multiply this result by $\sqrt{2}$, the scaling factor from one-dimensional to two-dimensional RMS for a symmetric system. Related tensor network simulations have been performed in \cite{hubig_evaulation_2020}.

\subsection*{Spinon-Holon Model}
We predict the spin correlations near the system center and the RMS hole distance using a spinon-holon model, described extensively in \cite{grusdt_parton_2018,grusdt_microscopic_2019, bohrdt_dynamical_2020}. The wave function separates into a spinon part and a string-holon part. For the spinon we follow the procedure described in \cite{bohrdt_dynamical_2020}: We equate the spinon dispersion relation with the zero-temperature magnetic polaron dispersion. The latter can be fitted by the functional form $\omega_{\rm sp}(\vec{k})= A [\cos(2k_x)+\cos(2k_y)] + B [\cos(k_x+k_y) + \cos(k_x-k_y)]$, with parameters $A=0.25 J$ and $B=0.36 J$ at $t/J=2$ \cite{martinez_spin_1991}. These values change very weakly when $t/J$ is increased, e.g. $A=0.28 J$ and $B=0.32 J$ for $t/J=3.5$. Using $\omega_{\rm sp}(\vec{k})$ and starting from a spinon localized on the central site, we simulate a quantum random walk. For the string-holon part, we follow \cite{grusdt_parton_2018} and solve the Schr\"odinger equation on the Bethe lattice in a linear potential, truncated at a depth $\ell_\text{trunc}$. We take the initial state fully localized on the root node, reducing the problem to a 1D problem on $\ell_\text{trunc}$ sites, which we solve by exact diagonalisation.

The parameters of the linear potential (the string tension and the $\ell=0$ offset) are computed from experimental correlations as indicated in \cite{grusdt_microscopic_2019}. For the $U/t=8.72$ data, we use the $\tau=0$ dataset, excluding the $3\times 3$-site region surrounding the hole in the tension computation to eliminate the suppressed correlations around the hole. At $U/t=17.2$, we lack correlation measurements at $\tau=0$, so we use a separate dataset taken under experimental conditions but with no hole initialized in the center. This likely overestimates the offset, as it ignores the suppression of correlations around the hole, but the simulation results are quite insensitive to the offset, and therefore still reliable. We find string tensions of $0.217(7)t$ ($0.101(4)t$)
and offsets of $0.002(3)t$ ($0.046(2)t$) at $U/t=8.72$ ($17.2$). These values justify using truncation lengths of $\ell_\text{trunc}=33$ ($100$) for $U/t=8.72$ ($17.2$), guaranteeing that $>99\%$ of the wave function remains with $\ell<\ell_\text{trunc}$ at all times.

From the separable wave function, we compute probability distributions $p_s(\mathbf{r}_s, \tau)$ and $p_h(\ell, \tau)$ for the spinon position and the path length on the Bethe lattice, and from the latter compute the probability distribution $p_{h,\text{rel}}(\mathbf{r}_h-\mathbf{r}_s, \tau)$ of the hole position relative to the spinon. Summing the RMSes of $p_s$ and $p_{h,\text{rel}}$ in quadrature yields the hole RMS distance. We use a $21\times 21$- ($23\times 23$-) site region of interest (ROI) for the spinon (hole) in this computation. This guarantees $\sim 95\%$ ($>99\%$) of the wave function remains within the ROI.

To compute spin correlations, we randomly sample nodes on the Bethe lattice at every depth $\ell<\ell_\text{trunc}$ and use each node to shuffle the values in the binarized atom matrices measured at $\tau=0$ by moving the initial central matrix value along the path corresponding to the Bethe lattice node. We only use those matrices with $0$ measured in the central site. At every $\ell<\ell_\text{trunc}$, we randomly sample a set of these shuffled matrices, and compute $pp$ and $p$ averages (defined in \cite{parsons_site-resolved_2016}), denoted $\langle pp\rangle^{\ell}$ and $\langle p\rangle^{\ell}$, within each set. We then use the separable wave function to predict these averages in the spinon-holon theory, e.g. as $\langle pp\rangle_{\mathbf{r},\mathbf{r}+\mathbf{d}}^\text{pred}(\tau)=\sum_{\mathbf{r}_s, \ell}p_h(\ell, \tau)p_s(\mathbf{r}_s, \tau)\langle pp\rangle^{\ell}_{\mathbf{r}-\mathbf{r}_s, \mathbf{r}+\mathbf{d}-\mathbf{r}_s}$. Once these spinon-holon averages are computed, we assemble them into spin correlators as in \cite{parsons_site-resolved_2016}. For every $\ell<\ell_\text{trunc}$, we sample 100 paths on the Bethe lattice. Increasing these numbers does not change the results. We assign errors to the $p$ and $pp$ averages by bootstrapping on the sampled ensembles at each depth $\ell$, and to $p_h$ and $p_s$ by varying their input parameters (string tension, $J$, etc.) by one standard deviation. The error bands on the predicted correlators in Fig.~5 are then obtained by linear error propagation. We account for the finite spinon escape probability in the predicted correlators by mixing in half the equilibrium correlator value with an inflated error bar; see next section. For the `Bethe' predictions in Fig.~5, we repeat this procedure with $J=0$ and a truncation depth of $\ell_\text{trunc}=100$.

\subsection*{Extended Discussion on Spinon-Holon Model}

We generate predictions for the spin correlations near the system center and the RMS hole distance as a function of time by applying a spinon-holon model of the dynamics to experimental data from $\tau=0$. This model has been described extensively in \cite{grusdt_parton_2018, grusdt_microscopic_2019, bohrdt_dynamical_2020}, so we will only outline it here. This theory predicts the correlation curves shown in Fig.~5 and the RMS distance curves in Extended Data Fig.~\ref{fig:RMS_LST}.

\if 1\mode
\placefigure[!htbp]{ef8}
\fi

The model assumes that the hole dynamics are much faster than the spin dynamics, so that the hole experiences an effectively static spin background (the \textit{frozen spin approximation}). Under this assumption, the hole must displace the spins that lie on the trajectory it takes across the lattice, leaving a trail of spins that are misaligned with the antiferromagnetic (AFM) background. This trail, called a \textit{geometric string}, extends from the hole to the site from which it was released, where the spin correlations are disturbed in a characteristic pattern called a \textit{spinon}. Spin dynamics are then introduced into this picture by passing from the strict frozen spin approximation to a Born-Oppenheimer approximation, in which spin dynamics are merely assumed to be adiabatic relative to the hole motion. This allows the spinon to hop on the lattice via spin exchange. 

The disturbance of the adjacent spin correlations $C_\mathbf{r}(|\textbf{d}| = 1)$ along the geometric string associates a magnetic energy to the string. In a system with local AFM correlations, this energy is usually positive (because, e.g., negative adjacent correlations get replaced with positive diagonal correlations $C_\mathbf{r}(|\textbf{d}| = \sqrt{2})$) and usually increases with string length, so that the hole is bound to the spinon by a confining \textit{string potential}. The spinon-hole system is thus a composite object, the motion of which is controlled by the slow spinon motion. Since the spinon carries spin and no charge, while the hole carries charge and no spin, the hole is sometimes called a \textit{chargon} or \textit{holon}.

To implement the Born-Oppenheimer dynamics in our simulation, we assume that the system begins and remains in a separable state, $\psi(\tau)=\psi_s(\tau)\psi_h(\tau)$, where $\psi_s$ and $\psi_h$ are spinon and string-hole wave functions (that is, $\psi_h$ describes not just the hole, but also the string that links it to the spinon). The spinon wave function is obtained as in \cite{bohrdt_dynamical_2020}
by initializing the wave function fully localized at the origin, and then time-evolving under the assumption that the spinon dispersion is given by the zero-temperature magnetic polaron dispersion from \cite{martinez_spin_1991}. This procedure is motivated by the observation that the spinon motion controls the motion of the bound spinon-holon pair (i.e., the magnetic polaron), so that the two objects should have equal dispersion relations. Since the magnetic polaron dispersion from \cite{martinez_spin_1991} is well-fit near $t/J=2$ by a tight-binding dispersion $\omega(\mathbf{k})=A[\cos(2k_x)+\cos(2k_y)]+B[\cos(k_x+k_y)+\cos(k_x-k_y)]$ with $A=0.25J$ and $B=0.36J$, this time-evolution procedure is equivalent to evolving under a tight-binding Hamiltonian with diagonal and straight next-nearest-neighbor hopping, and therefore tractable by exact diagonalisation.

To compute the string-hole wave function, we first make the Bethe-lattice approximation to the string-hole Hilbert space (as in \cite{grusdt_parton_2018}). That is, we assume that all of the trajectories the hole can take on the square lattice that do not contain self-retracing components correspond to different states in the string-hole Hilbert space. This approximation, which is also made for the `Bethe' curve in Fig.~3, is motivated by the observation that, due to the AFM background, most of the trajectories the hole can take result in different spin configurations and are therefore distinguishable.

Solving for the string-hole wave function then reduces to solving for the dynamics of an initially localized wavepacket on the Bethe lattice in the string potential. We make the approximation of \textit{linear string theory} (LST), in which the magnetic energy of a string is assumed to increase linearly with its length. Specifically, we assume that the potential energy of a string $\Sigma$ is $V(\Sigma)=\left(\frac{dE_{\Sigma}}{d\ell}\right)\ell(\Sigma)-g\delta_{0\ell(\Sigma)}$, where $\ell(\Sigma)$ is the length of $\Sigma$, the \textit{string tension} $dE_{\Sigma}/d\ell$ gives the increase in string energy per unit length, and $g$ is an offset that captures the deviation of $(V|_{\ell(\Sigma)=1}-V|_{\ell(\Sigma)=0})$ from the string tension. Roughly speaking, this deviation occurs because the first step the hole takes away from its initial site breaks more AFM bonds than do its later steps.

This problem is solved in \cite{grusdt_parton_2018}, where it is shown that the eigenstates of the problem correspond to vibrational and rotational excitations of the string. Rotationally excited states have nontrivial phase windings as one circles the root node of the Bethe lattice. We assume that our initial state is perfectly localized on the root node of the Bethe lattice, that is, $\psi_h(\Sigma, \tau=0)=\delta_{0\ell(\Sigma)}$. This wave function has no overlap with the rotationally excited states, due to its isotropy, which in turn implies that $\psi_h(\Sigma, \tau)$ is isotropic at all times -- that is, $\psi_h(\Sigma, \tau)$ depends only on $\ell(\Sigma)$ and $\tau$.

Together with the LST potential, this reduces the string-hole problem to a problem on a half-infinite one-dimensional chain of sites, labeled by $\ell=0, 1, 2\ldots$, where site $\ell$ corresponds to the $\ell^\text{th}$ layer of the Bethe lattice. The initial wave function is $\psi_{1D}(\ell, \tau=0)=\delta_{0\ell}$. It should be noted that this one-dimensional wave function is rescaled from the wave function on the Bethe lattice by the multiplicity of a layer of the Bethe lattice, that is, $\psi_h(\Sigma, \tau)=\psi_{1D}(\ell(\Sigma), \tau)/\sqrt{4\cdot 3^{\ell(\Sigma)-1}}$. The one-dimensional Hamiltonian is a sum of $V(\ell)=\left(\frac{dE_{\ell}}{d\ell}\right)\ell-g\delta_{0\ell}$ and a hopping term $T_{\ell\ell'}$, originating from the hopping term in the Hubbard model, which connects different sites of the chain. As shown in \cite{grusdt_parton_2018},
% \if 0\mode\begin{linenomath}\fi
\begin{equation*}
T_{\ell\ell'}=
\begin{cases} -2t, & \ell=0, \ell'=1\textrm{ or }\ell=1, \ell'=0\\
-t\sqrt{3}, & |\ell-\ell'|=1, \ell>0, \ell'>0 \\
0, & \textrm{else}
\end{cases}
\end{equation*}
% \if 0\mode\end{linenomath}\fi
The factors of $2$ and $\sqrt{3}$ are due to the connectivity between adjacent layers of the Bethe lattice.

We solve the string-hole problem by diagonalising the Hamiltonian $H^{1D}_{\ell\ell'}=T_{\ell\ell'}+V(\ell)\delta_{\ell\ell'}$, decomposing the initial wavepacket into its eigenstates, and time-evolving. We truncate the Hamiltonian at $\ell=33$ for the data in Fig.~5 and $\ell=100$ for the data in Fig.~\ref{fig:RMS_LST} (these truncation $\ell$ values are exclusive endpoints). These truncation lengths are sufficient to retain nearly all of the wave function's probability amplitude in the truncation region at all times in the simulations ($>99\%$ in Fig.~5, and $100\%$ to within machine precision in Fig.~\ref{fig:RMS_LST}). It should be noted that the retained probability for a given truncation length depends on the string tension; our truncation lengths were chosen based on our measured string tensions.

Together with the spinon wavefunction, this yields the full Born-Oppenheimer wave function $\psi(\mathbf{r}_s, \Sigma, \tau)=\psi_s(\mathbf{r}_s, \tau)\psi_{1D}(\ell(\Sigma), \tau)/\sqrt{4\cdot 3^{\ell(\Sigma)-1}}$, where $\mathbf{r}_s$ is the spinon position and $\Sigma$ is a string (or equivalently, a node on the Bethe lattice). With these wave functions, we may also immediately calculate $p_s(\mathbf{r}_s, \tau)$ and $p_h(\ell, \tau)$, respectively the probabilities of finding the spinon at $\mathbf{r}_s$ and the holon at depth $\ell$ on the Bethe lattice at time $\tau$. This wave function is used to generate the spinon-holon curves in Fig.~5 and Fig.~\ref{fig:RMS_LST} as follows.

To compute the RMS hole distance, we first convert the string-hole wave function to a probability distribution $p_{h,\text{rel}}$ for $\mathbf{r}_h-\mathbf{r}_s$, where $\mathbf{r}_h$ is the hole position. This is possible because every string $\Sigma$ can be mapped (by following the path $\Sigma$ takes on the square lattice) to a relative position $\mathbf{r}_h-\mathbf{r}_s$, so that converting $p_h(\ell, \tau)$ to $p_{h,\text{rel}}(\mathbf{r}_h-\mathbf{r}_s, \tau)$ only requires solving the combinatorial problem of how many paths at each layer $\ell$ of the Bethe lattice arrive at each relative position $\mathbf{r}_h-\mathbf{r}_s$. The RMS distance of the hole is then given by $d_\text{RMS}(\tau)^2=\sum_{\mathbf{r}_h, \mathbf{r}_s}p_s(\mathbf{r}_s, \tau)p_{h,\text{rel}}(\mathbf{r}_h-\mathbf{r}_s, t)\mathbf{r}_h^2$, which can be computed either directly or by summing the RMS distances of $p_{h,\text{rel}}$ and $p_s$ in quadrature. 

To compute the spin correlations, we shuffle the experimental snapshots measured at $\tau=0$ (just before the hole is released) according to the Born-Oppenheimer wave function, and compute the spin correlations of these shuffled ensembles of snapshots as we do with normal experimental data (see \cite{parsons_site-resolved_2016}). More specifically, for every string length $\ell$ up to the truncation length, we sample a random set $\tau_{\ell}$ of strings (uniformly with replacement) from all possible strings of length $\ell$. For every string $\Sigma$ in $\tau_{\ell}$ and every experimental snapshot $M$ at $\tau=0$, we generate the shuffled snapshot $M_{\Sigma}$ that results from rearranging the values in the snapshot as the hole would the spin background in the frozen spin approximation.
We compute $\mathcal{M}_{\ell}$, the set of all $M_{\Sigma}$ for $\Sigma\in\tau_{\ell}$.
% We then randomly sample sets $\mathcal{M}_{\ell}$ (uniformly and with replacement) from the set of all $M_{\Sigma}$ generated by all $\Sigma\in\tau_{\ell}$ (we also enforce the constraint that each $\mathcal{M}_{\ell}$ has an equal number of snapshots with no spin removal and with each type of spin removal).
Then, treating each $\mathcal{M}_{\ell}$ as an independent ensemble of snapshots, we compute the $pp$ and $p$ averages (defined in \cite{parsons_site-resolved_2016}) in each $\mathcal{M}_{\ell}$ across a $23\times 23$-site ROI centered on the initial hole site. Denote these averages as $\langle pp\rangle^{\ell}_{\mathbf{r}, \mathbf{r}+\mathbf{d}}$ and $\langle p\rangle^{\ell}_{\mathbf{r}}$. We then compute the corresponding averages in the spinon-holon theory as $\langle pp\rangle^\text{pred}_{\mathbf{r}, \mathbf{r}+\mathbf{d}}(\tau)=\sum_{\mathbf{r}_s, \ell}p_h(\ell, \tau)p_s(\mathbf{r}_s, \tau)\langle pp\rangle^{\ell}_{\mathbf{r}-\mathbf{r}_s, \mathbf{r}+\mathbf{d}-\mathbf{r}_s}$
and
$\langle p\rangle^\text{pred}_{\mathbf{r}}(\tau)=\sum_{\mathbf{r}_s, \ell}p_h(\ell, \tau)p_s(\mathbf{r}_s, \tau)\langle p\rangle^{\ell}_{\mathbf{r}-\mathbf{r}_s}$. We then assemble the predicted $p$ and $pp$ averages at each time into the sign-corrected, polarization-corrected spin correlator $C_{\mathbf{r}}(\mathbf{d})$ defined in \cite{parsons_site-resolved_2016}.  For the `Bethe' curves in Fig.~5, we repeat this procedure with $J=0$ and $\ell_\text{trunc}=100$.

This procedure is equivalent to associating the configuration with spinon position $\mathbf{r}_s$ and string $\Sigma$ with the snapshots $M_{\Sigma, \mathbf{r}_s}$ that result from translating $M_{\Sigma}$ by $\mathbf{r}_s$. This is the most natural way to handle the hopping of the spinon in our simulation, as it is based on the translational invariance of the system at $\tau=0$ away from the initial hole site. The functions $C_{\mathbf{r}}(\mathbf{d})$ at time $\tau$ are then computed to equal the correlations that would result from the ensemble of snapshots $\{M_{\Sigma, \mathbf{r}_s}\}$, weighted by the probability distribution $|\psi(\mathbf{r}_s, \Sigma, t)|^2$. Random sampling is only introduced to the procedure to deal with the exponentially large number of paths for longer strings. For the curves plotted in Fig.~5, each $\tau_{\ell}$ had size $100$. This value is sufficiently large that the resulting curves do not appreciably change upon repeating the random sampling or increasing the number of samples.

Errors are calculated as follows. For the $p$ and $pp$ averages, we use bootstrap sampling across all experimental snapshots $\mathcal{M}$ to generate error bars -- we must bootstrap across $\mathcal{M}$ rather than $M_{\Sigma}$ to capture the strong covariances between data generated from the same experimental snapshot. For the probability distributions $p_h(\ell, \tau)$ and $p_s(\mathbf{r}_s, \tau)$, we first increase, then decrease the experimentally measured input parameters (string tension, offset, $J$, and $t$) by one standard deviation, and recalculate the probability distributions to generate distributions $p_h^{\pm}(\ell, \tau)$ and $p_s^{\pm}(\mathbf{r}_s, \tau)$. The error bars we assign to $p_h(\ell, \tau)$ and $p_s(\mathbf{r}_s, \tau)$ are then $\frac{1}{2}|p_h^{+}(\ell, \tau)-p_h^{-}(\ell, \tau)|$ and $\frac{1}{2}|p_s^{+}(\mathbf{r}_s, \tau)-p_s^{-}(\mathbf{r}_s, \tau)|$. We then propagate errors through to the quantities $\langle pp\rangle^\text{pred}_{\mathbf{r}, \mathbf{r}+\mathbf{d}}(\tau)$ and $\langle p\rangle^\text{pred}_{\mathbf{r}, \mathbf{r}+\mathbf{d}}(\tau)$ by standard linear error propagation.

Finally, after assembling the predicted $p$ and $pp$ averages into the correlators $C_{\mathbf{r}}(\mathbf{d})$, we account for error induced by the finite probability $p_\text{esc}(\tau)=1-\sum_{\mathbf{r}_s}p_s(\mathbf{r}_s, \tau)$ that the spinon escapes its ROI. We do this by adding a quantity $p_\text{esc}C_\text{eq}(\mathbf{d})$ onto the predicted $C_{\mathbf{r}}(\mathbf{d})$, where $C_\text{eq}(\mathbf{d})$ is the background value of $C_{\mathbf{r}}(\mathbf{d})$ at $\tau=0$ (obtained by averaging $C_{\mathbf{r}}(\mathbf{d})$ with $\mathbf{r}$ in a $23\times 23$ ROI with a $3\times 3$ ROI about the origin excluded to avoid the reduced correlations around the hole). The motivation for this procedure is to allow `missing' snapshots, where the spinon escapes its ROI, to contribute a random value between $0$ and $C_\text{eq}(\mathbf{d})$ to the predicted correlator.

\begin{dfigure}{ef8}
    \centering
    \includegraphics[width=\figwidth]{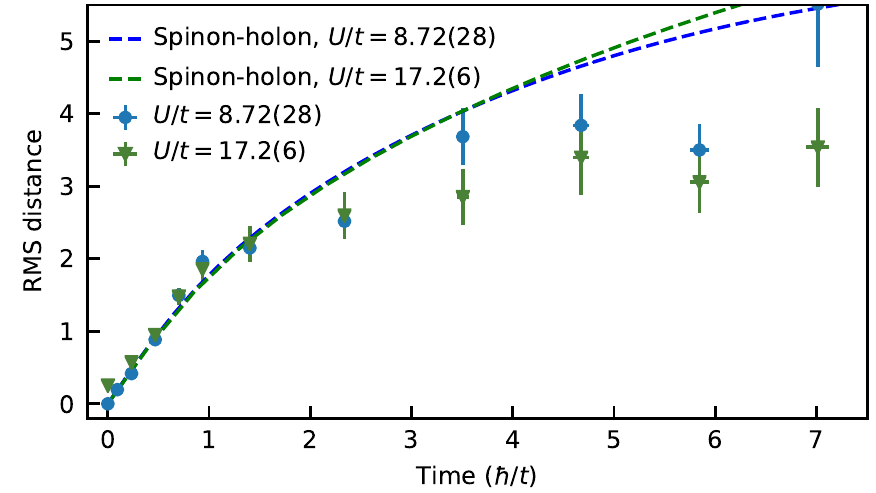}
    \caption{RMS hole distance as measured in experiment (points) and predicted by spinon-holon model (dashed curves).}
    \label{fig:RMS_LST}
\end{dfigure}

\begin{dfigure}{uniformity}
    \centering
    \includegraphics[width=\figwidth]{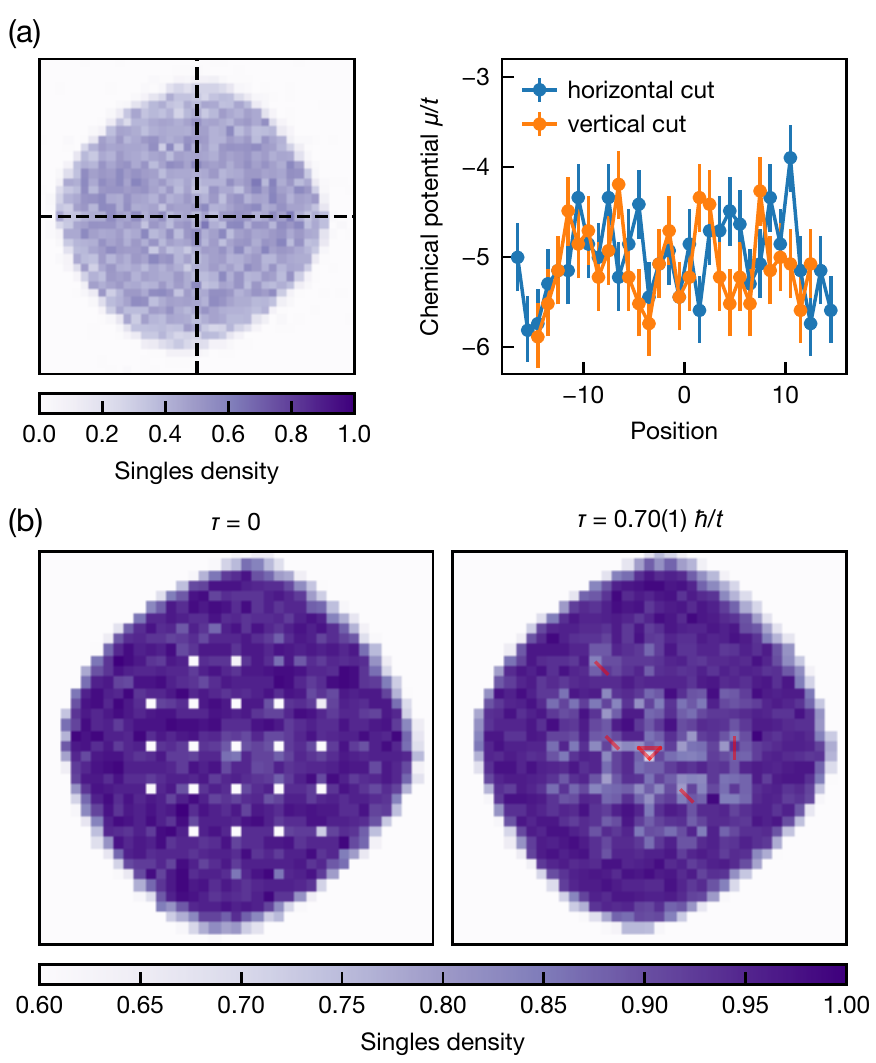}
    \caption{(\textbf{a}) Left: Site-resolved singles density in the metallic regime, averaged over 90 realizations. Right: Local chemical potential along the dashed lines, inferred from NLCE \cite{khatami_thermodynamics_2011} and defined with respect to the chemical potential at half-filling. Within local density approximation, this corresponds to a site-to-site RMS deviation of the underlying potential of about $0.4 t$ across the filled region. (\textbf{b}) Singles density in the Mott-insulating regime upon pinning about 20 holes on a square grid, and after a release time of $\tau = 0.7(1) \hbar/t$. The density maps are averaged over 118 and 164 realizations respectively. Nearest neighbors to the same initial hole with statistically significant density differences (with a 95\% confidence level) are highlighted by red links at $\tau = 0.7(1) \hbar/t$.}
    \label{fig:uniformity}
\end{dfigure}

\begin{dfigure*}{initial_density}
    \centering
    \includegraphics[width=\textwidth]{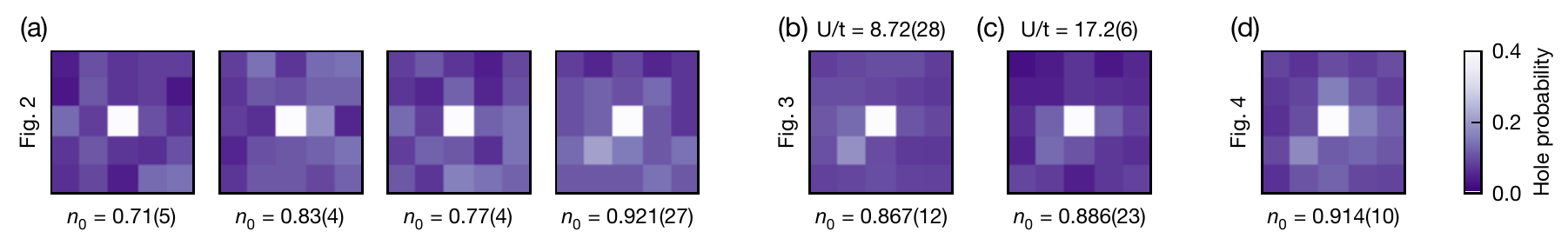}
    \caption{\textbf{Initial hole probability.} (\textbf{a}) Four-hole dataset used in Figs.~2 and 3, averaged over 100 realizations. (\textbf{b}) Single-hole dataset used in Fig.~3, at U/t = 8.72(28) (762 realizations) and (\textbf{c}) U/t = 17.2(6) (193 realizations). (\textbf{d}) Single-hole dataset used in Figs.~4 and 5 (764 realizations).}
    \label{fig:initial_density}
\end{dfigure*}

\begin{dfigure*}{density_full}
    \centering
    \includegraphics[width=0.8\textwidth]{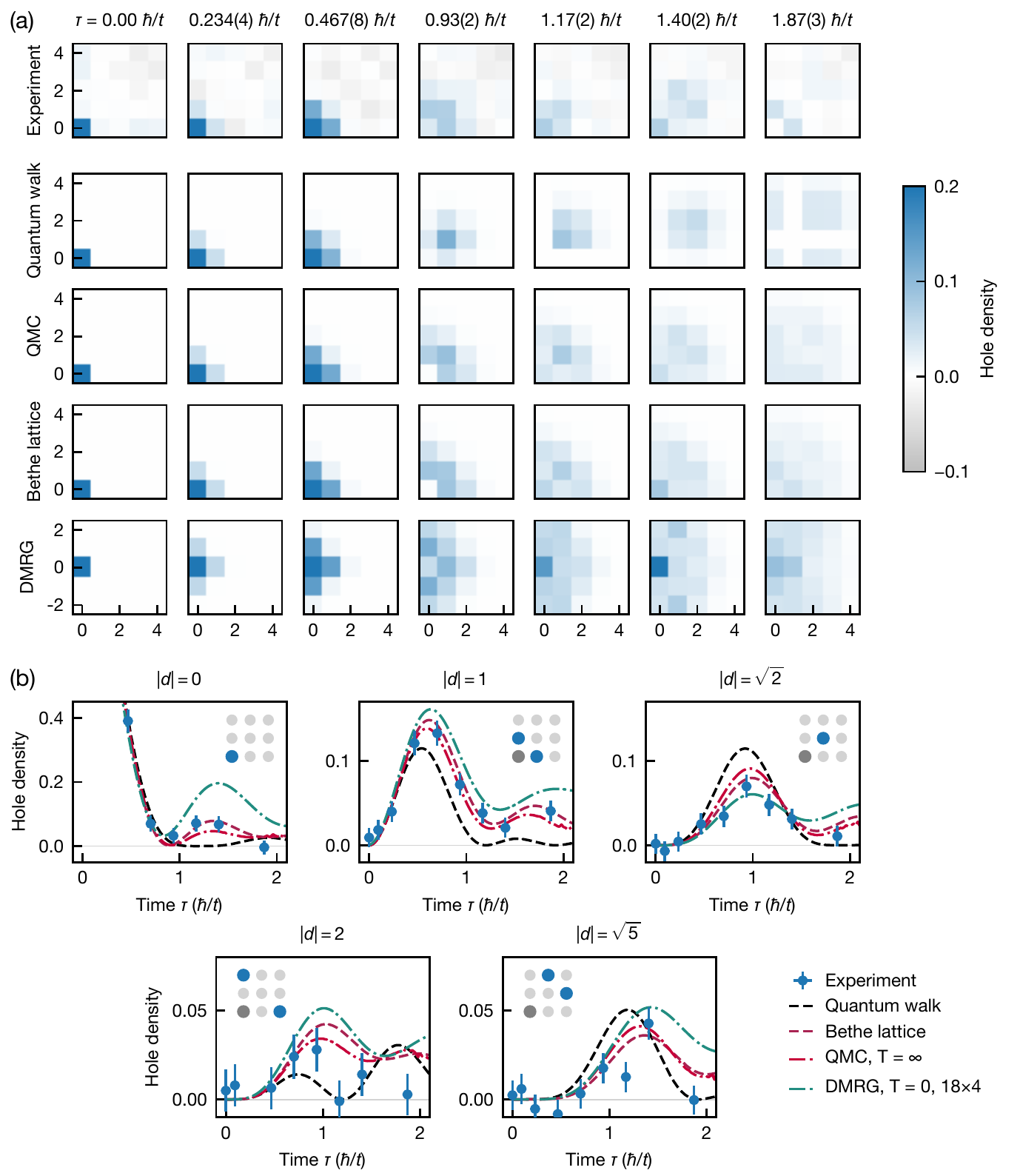}
    \caption{\textbf{Full comparison between experimental and simulated densities.} (\textbf{a}) Two-dimensional hole density around the origin, clipped to $5 \times 5$ quadrants. The experimental data is averaged over horizontal, vertical and diagonal reflections. The DMRG simulation is performed on a $18 \times 4$ system with periodic boundary conditions along the short vertical direction; the rows at $d_y = \pm 2$ are equivalent and are duplicated for clarity. (\textbf{b}) Hole density averaged over sites at distances $|\textbf{d}| = 0, 1, \sqrt{2}, 2$ and $\sqrt{5}$.}
    \label{fig:density_full}
\end{dfigure*}

\begin{dfigure*}{fit_total_in_mask}
    \centering
    \includegraphics[width=\textwidth]{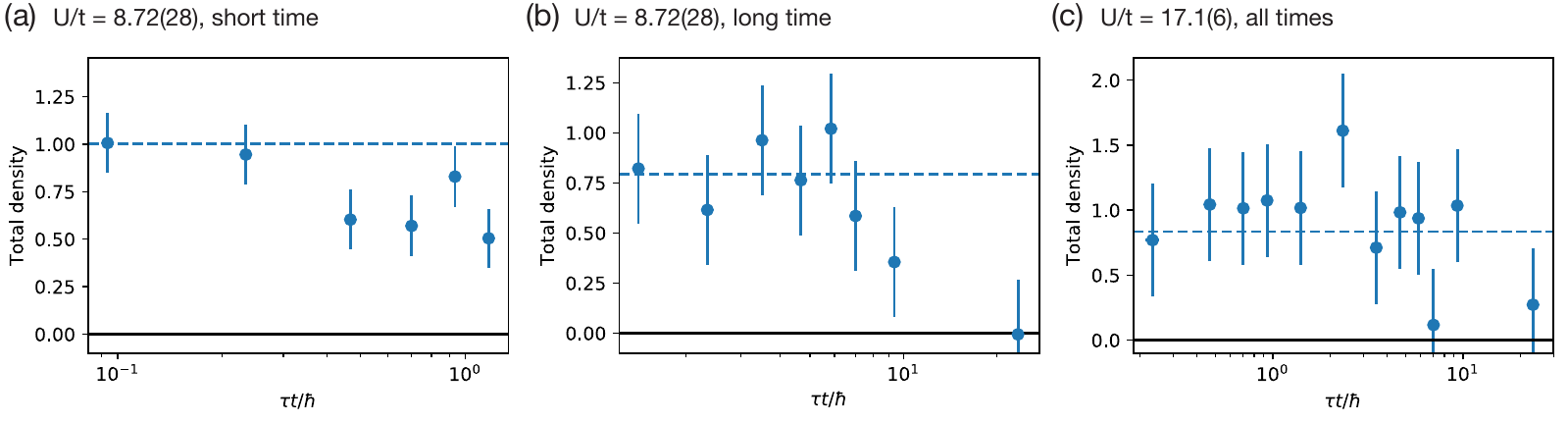}
    \caption{Plot of the total hole density inside the analysis window. For both values of U/t, we find that the hole density is consistent with $0$ at the last measured time of $\tau t/\hbar=23.4$. The blue dashed line is the hole density at $t=0$.}
    \label{fig:fit_total_in_mask}
\end{dfigure*}

\begin{dfigure*}{fit_red_chisq}
    \centering
    \includegraphics[width=0.9\textwidth]{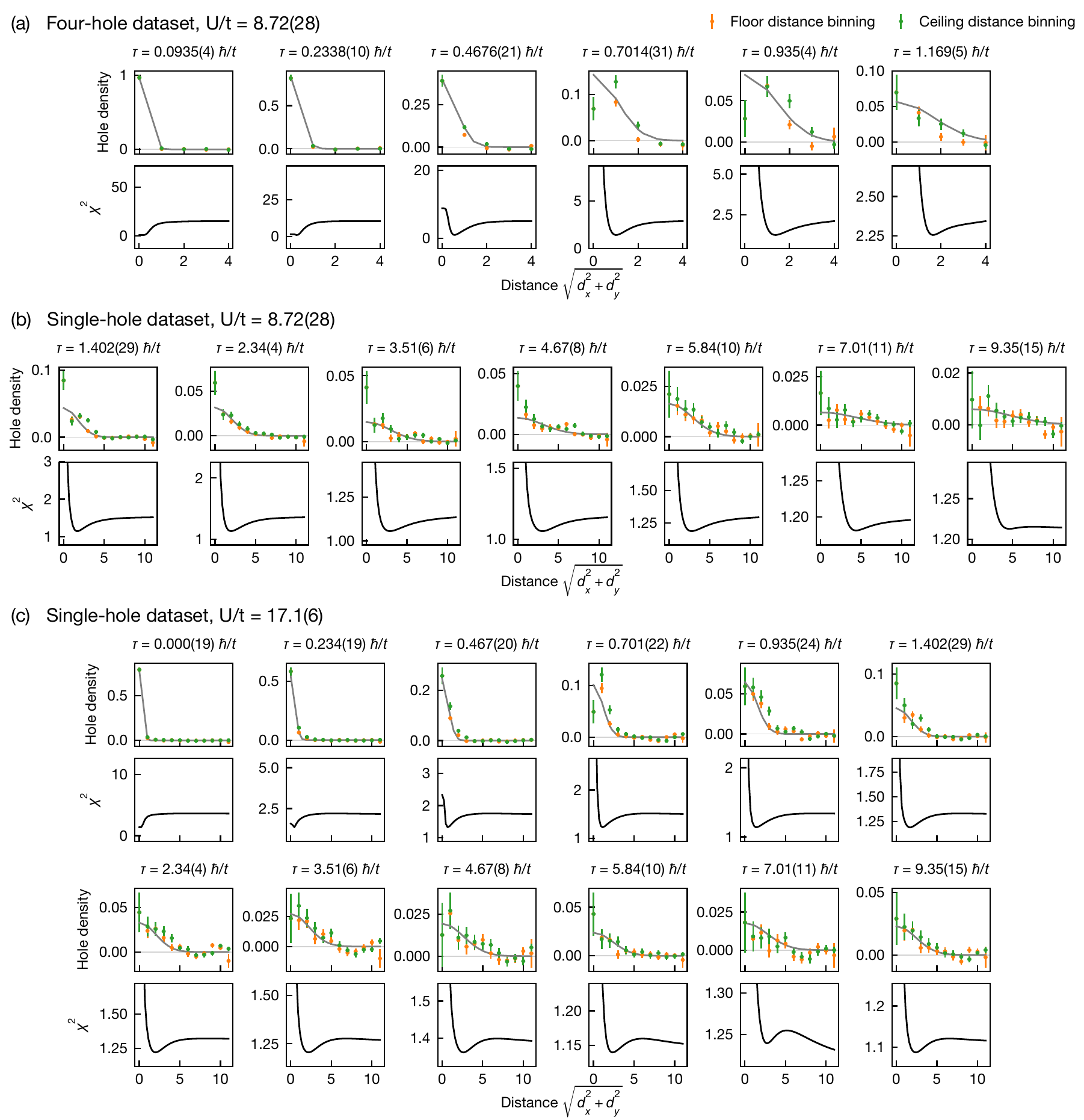}
    \caption{\textbf{Gaussian fits to experimental hole density distributions.} Experimental (yellow and green points) hole density and its Gaussian fit (gray line) as a function of Euclidian distance, and resulting reduced $\chi^2$ test statistics (black line) as a function of the width of the Gaussian function. The number of experimental points displayed is reduced for clarity by binning them according to the floor or ceiling of their distance. The last measured time for each value of $U/t$ is not included in the main text due to the sensitivity of the fit to the exact window size used.}
    \label{fig:fit_red_chisq}
\end{dfigure*}

\begin{dfigure}{fit_investigation}
    \centering
    \includegraphics[width=\figwidth]{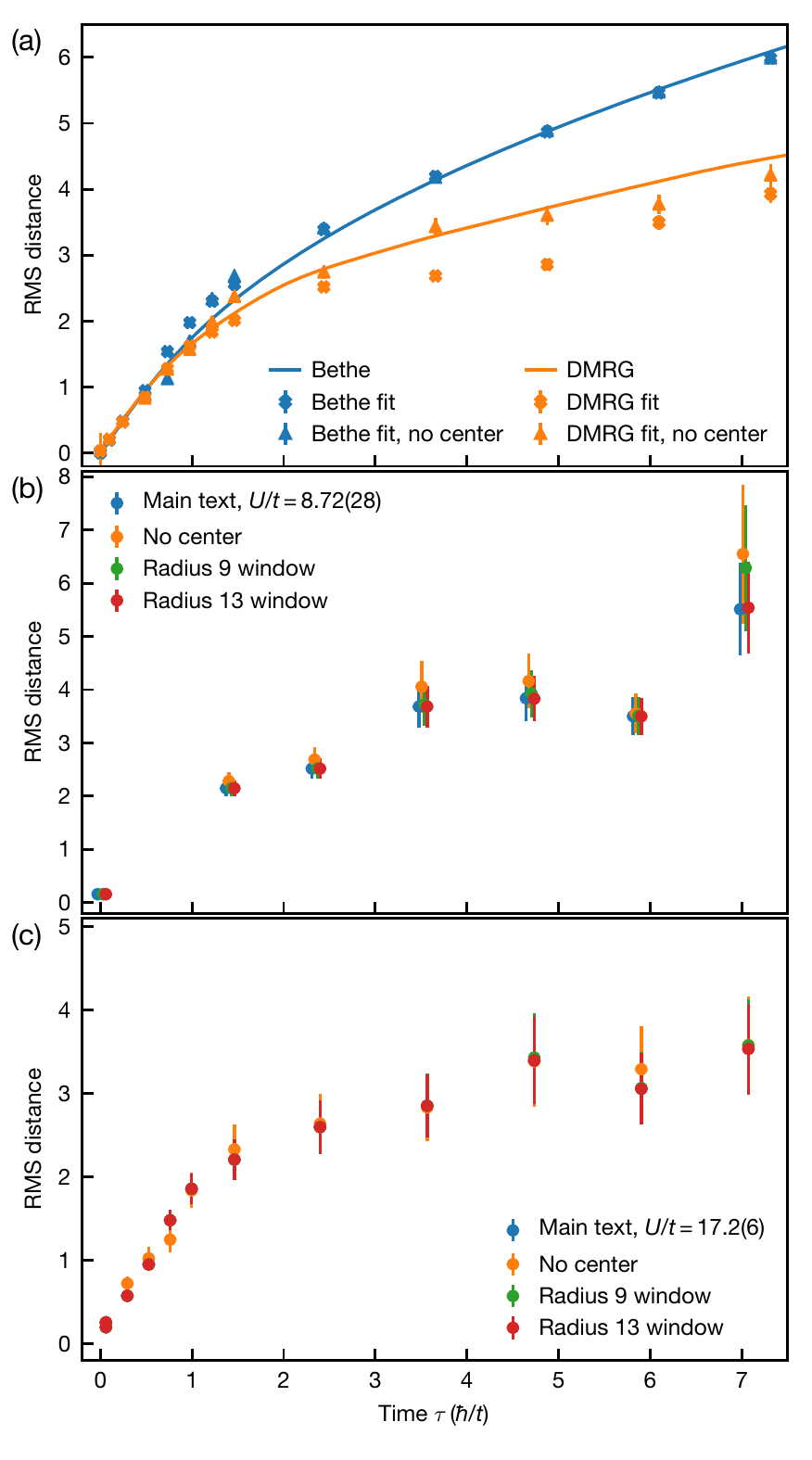}
    \caption{\textbf{Suitability of the Gaussian fits.} (\textbf{a}) RMS extracted from Gaussian fits including and excluding the center point (rung) for the Bethe-lattice model and TD-DMRG simulations, along with the RMS calculated from the bare distributions. Excluding the center point improves the TD-DMRG fit, which we believe is due to a finite-size effect. (\textbf{b, c}) RMS extracted from Gaussian fits on the experimental data for different-sized fitting windows and with and without the center point. The fit results are not dramatically affected by these changes.}
    \label{fig:fit_investigation}
\end{dfigure}

\begin{dfigure*}{bondmap}
    \centering
    \includegraphics[width=\textwidth]{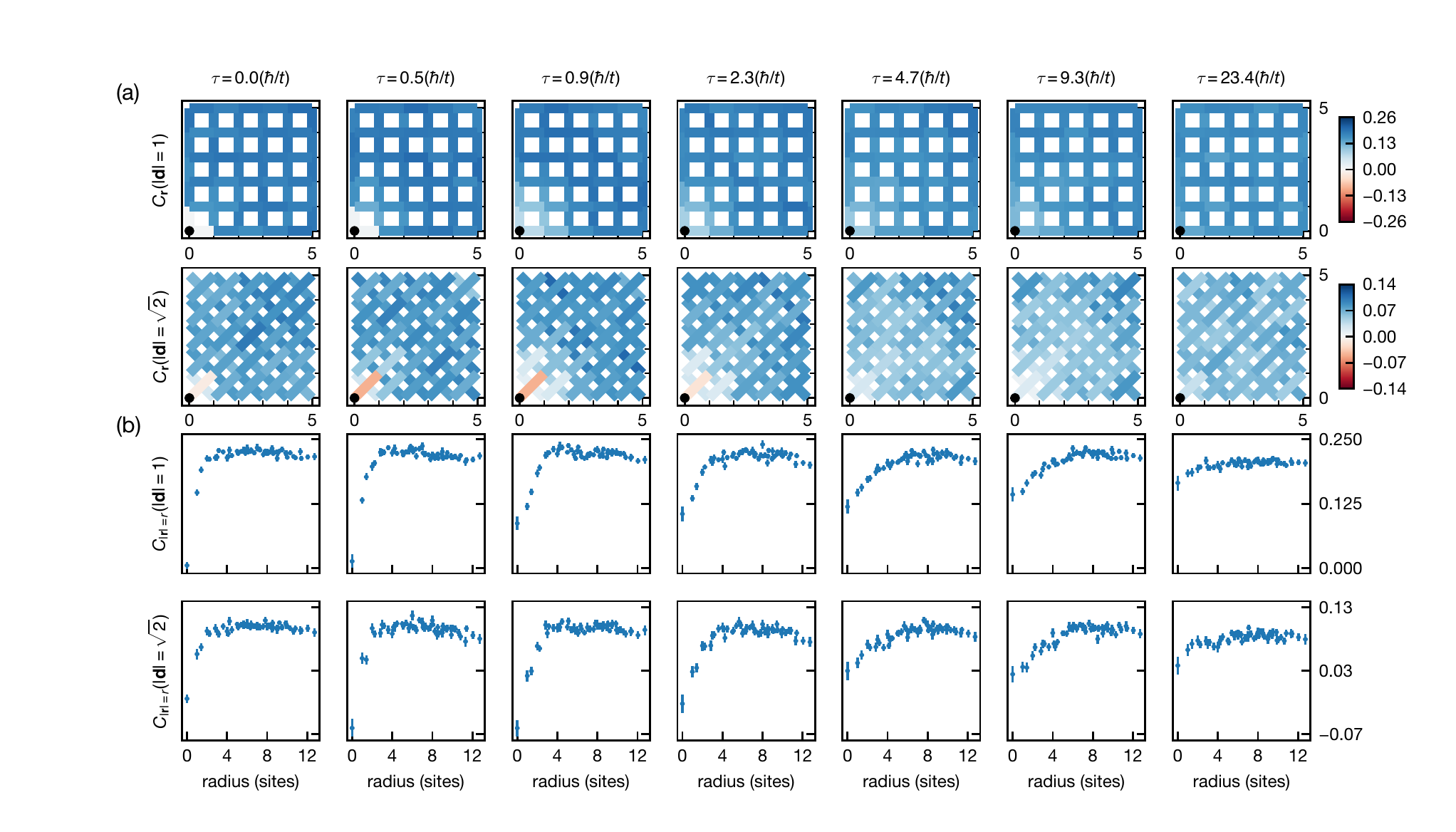}
    \caption{\textbf{Spatial dependence of the adjacent and diagonal spin correlations.} (\textbf{a}) Bond-specific adjacent and diagonal spin correlators in a $11\times 11$-site ROI around the hole as a function of time, symmetrised across reflections. The correlators are sign-corrected as in Figs.~4 and 5, and are predominantly antiferromagnetic for adjacent correlators (first row) and ferromagnetic for diagonal correlators (second row). (\textbf{b}) Radial averages of the same correlators, now from a $21\times 21$-site ROI.}
    \label{fig:bondmap}
\end{dfigure*}

%%%%%%%%%%%%%%%%%%%%%%%%%%%%%%%%%%%%%%%%%%%%%%%%%%%%%%%%%%%%%%%%%%%%%%%%%%%%%%